\documentclass[12pt]{article}

\usepackage{amsmath,amsfonts,amssymb}

\def\bea{\begin{eqnarray}}
\def\eea{\end{eqnarray}}
\def\be{\begin{equation}}
\def\ee{\end{equation}}
\def\ba{\begin{array}}
\def\ea{\end{array}}
\def\nn{\nonumber}

\font\tenrsfs=rsfs10
\font\sevenrsfs=rsfs7
\font\fiversfs=rsfs5
\newfam\rsfsfam
\textfont\rsfsfam=\tenrsfs
\scriptfont\rsfsfam=\sevenrsfs
\scriptscriptfont\rsfsfam=\fiversfs
\def\mathscr#1{{\fam\rsfsfam\relax#1}}

\renewcommand{\theequation}{\arabic{section}.\arabic{equation}}
\begin{document}

\begin{titlepage} 

\rightline{\footnotesize{CERN-PH-TH/2008-215}} \vspace{-0.2cm}
\rightline{\footnotesize{ZMP-HH/08-16}} \vspace{-0.2cm}

\begin{center}

\vskip 0.4 cm

\begin{center}
{\Large{ \bf No metastable de Sitter vacua in ${\cal N}\!=2$ \\[2mm] supergravity with only hypermultiplets}}
\end{center}

\vskip 1cm

{\large 
Marta G\'omez-Reino$^{a}$, Jan Louis$^{b,c}$ and Claudio A. Scrucca$^{d}$
}

\vskip 0.8cm

{\it
$^{a}$Theory Division, Physics Department, CERN, \\
CH-1211 Geneva 23, Switzerland\\
$^{b}$II. Institut f\"ur Theoretische Physik, Universit\"at Hamburg, \\
D-22761 Hamburg, Germany\\
$^{c}$Zentrum f\"ur Mathematische Physik,
Universit\"at Hamburg,\\
Bundesstrasse 55, D-20146 Hamburg, Germany\\
$^{d}$Institut de Th\'eorie des Ph\'enom\`enes Physiques, EPFL, \\
\mbox{CH-1015 Lausanne, Switzerland}\\
}

\vskip 0.8cm

\end{center}

\begin{abstract}

We study the stability of vacua with spontaneously broken supersymmetry in 
${\cal N}\!=2$ supergravity theories with only hypermultiplets. Focusing on the
projection of the scalar mass matrix along the sGoldstino directions, 
we are able to derive a universal upper bound on the lowest mass eigenvalue. 
This bound only depends on the gravitino mass and the cosmological constant, 
but not on the details of the quaternionic manifold spanned by the scalar fields. 
Comparing with the Breitenlohner-Freedman bound shows that metastability 
requires the cosmological constant to be smaller than a certain negative critical 
value. Therefore, only AdS vacua with a sufficiently negative cosmological constant 
can be stable, while Minkowski and dS vacua necessarily have a tachyonic direction.

\end{abstract}

\bigskip

\end{titlepage}

\newpage

\section{Introduction} \setcounter{equation}{0}

A crucial issue in string theory is to identify  a mechanism for
supersymmetry breaking which, at the same time, keeps the 
cosmological constant small, as current experimental
observations suggest the existence of a tiny positive cosmological 
constant (dark energy) driving the expansion of the universe that 
we observe today. This has motivated the search for four-dimensional 
de Sitter (dS) vacua in string theory.
One possible approach to this problem is to stay within the low-energy 
effective four-dimensional supergravity description and first determine 
the conditions under which a metastable vacuum exhibiting spontaneous 
supersymmetry breaking with a reasonably small cosmological constant 
can possibly arise. One may then similarly ask under which conditions it 
is possible to realize slow-roll inflation in such a setup. While finding the 
answers to these questions may not be sufficient for understanding the 
vacuum selection mechanism within string theory, it would certainly be 
a useful guideline for model building.

Arranging for metastable dS vacua in generic supersymmetric theories
turns out to be surprisingly difficult. One of the reasons is that these vacua 
necessarily break supersymmetry spontaneously and hence supersymmetry 
does not guarantee the stability of the ground state. Actually, in refs.~\cite{GRS,CGGLPS} 
a necessary condition for the existence of metastable dS vacua within generic 
${\cal N}\!=1$ supergravity theories was identified.\footnote{See also \cite{DD} 
for an analysis with similar spirit applied to the idea of landscape of vacua.}
The crucial physical ingredient exploited in these analyses is the fact 
that in the scalar field space the most dangerous directions for metastability 
are the ones corresponding to the sGoldstinos, the supersymmetric partners of the Goldstino. 
While all the other multiplets can be made arbitrarily massive by suitably  
tuning the superpotential, the Goldstino multiplet is only allowed to have
mass splittings induced by supersymmetry breaking. Thus the requirement 
for the sGoldstino square mass to satisfy the metastability bound (namely 
being positive in dS space and within the negative Breitenlohner-Freedman (BF)
bound \cite{BF} in anti de Sitter (AdS) space) is independent of the superpotential 
but instead poses a strong necessary condition on the curvature of the scalar
geometry. More precisely, along the sGoldstino direction the sectional curvature 
of the K\"ahler manifold spanned by the scalar fields has to have a limited size. 
Since the sGoldstino direction is determined by the superpotential 
this in turn poses also a constraint on the superpotential. 

The aim of this paper is to pursue a similar study for ${\cal N}\!=2$ supergravity theories. 
The motivation for doing this is two-fold: Firstly, the scalar field space of ${\cal N}\!=2$ 
supergravity is not a special case of the ${\cal N}\!=1$ field space. Also, the scalar potential 
in ${\cal N}\! = 2$ theories is fixed by a gauging of isometries, while the one of ${\cal N}\!=1$ 
theories is governed by an arbitrary superpotential. This makes the two analyses qualitatively 
different. Secondly, the hidden sector of string theory, where supersymmetry
is believed to be spontaneously broken, often displays such an extended supersymmetry. 
Therefore an analysis in extended supergravity theories seems to be more suitable to 
establish the relation with higher-dimensional theories.

As a first step of this program we will focus in this paper on the simple situation of ${\cal N}\!=2$ 
theories which involve only hypermultiplets. As we will see in the body of the paper these 
theories are in some sense the analogs of ${\cal N}\!=1$ theories with
only chiral multiplets.\footnote{However, they have the peculiarity of becoming trivial 
in the limit of rigid supersymmetry, where gravity is decoupled.}
The main result we find is that in ${\cal N}\!=2$ theories with only 
hypermultiplets metastability implies a negative upper bound on the cosmological constant 
and therefore dS vacua (as well as slow-roll inflation) are always excluded. A similar 
conclusion was also reached in the other particular situation of ${\cal N}\!=2$ theories involving 
only vector multiplets and Abelian gaugings, where metastability forces the cosmological constant 
to be negative \cite{DW,CKVDFDG}. The study of more general situations, involving both hyper 
and vector multiplets and/or non-Abelian gaugings, is left to a subsequent paper \cite{DGLS}. 
In this more general type of theories a richer variety of possibilities is expected to exist. In fact 
some particular examples of stable dS vacua have already been constructed in this context, 
for instance in refs.~\cite{FTV,O} exploiting non-Abelian gauge symmetries. 

The paper is organized as follows. In Section 2 we briefly review the results of refs.~\cite{GRS,CGGLPS}
using  a formalism that is tailored for the transition to ${\cal N}\!=2$ theories. In fact we slightly generalize 
the previous analyses in that we also derive a constraint for the existence of metastable AdS ground
states with spontaneously broken supersymmetry. In Section 3 we show that in ${\cal N}\!=2$ supergravities 
with only hypermultiplets no metastable dS ground states exist and derive a bound for the non-supersymmetric 
AdS vacua. Finally, in Section 4 we summarize our conclusion and give an outlook on future directions of 
investigation. For completeness, we record the computations of the supertrace sum rules on boson and 
fermion masses for ${\cal N}\!=1$ and ${\cal N}\!=2$ theories in Appendix~A. We also summarize our 
conventions for the curvature of real and complex manifolds in Appendix~B.

\section{{\cal N}=1 theories with chiral multiplets} \setcounter{equation}{0}

In order to prepare for the analysis in ${\cal N}\!=2$ supergravity, we shall start by
briefly reviewing the conditions for the existence of metastable vacua in spontaneously 
broken ${\cal N}\!=1$ supergravity. We follow our earlier papers \cite{GRS,CGGLPS}
but use a slightly modified formalism, which makes the transition to ${\cal N}\!=2$ 
theories somewhat more suggestive. Furthermore, in \cite{GRS,CGGLPS} we 
concentrated on finding dS vacua whereas in the following we extend the analysis 
to also include non-supersymmetric AdS vacua.

\subsection{Preliminaries}

Let us consider a generic ${\cal N}\!=1$ theory with $n$ chiral multiplets 
$\Phi^i$, containing complex scalar fields $\phi^i$ and chiral fermions $\chi^i$ \cite{CJSFGV}. 
This theory is described by  a superpotential $W$ and a K\"ahler potential $K$  which 
defines a K\"ahler-Hodge geometry with a metric for the scalar fields given by 
$g_{i \bar \jmath} = K_{i \bar \jmath}$. The theory has a $U(1)$ K\"ahler invariance 
which transforms $K \to K + f + \bar f$ and $W \to W e^{-f}$. The holonomy of the scalar 
manifold is contained in $U(1) \times U(n)$, where the $U(1)$ curvature form is 
identified with the K\"ahler form while the $U(n)$ curvature is arbitrary.

Instead of choosing a K\"ahler gauge and describing the theory in terms of the 
invariant function $G = K + {\rm log} |W|^2$, we will use instead a different formulation 
where this symmetry is kept manifest. For this purpose, it is useful to introduce the quantity 
\be
L \equiv e^{K/2} W \,.
\ee
$L$ transforms with weight $\tfrac 12$ under K\"ahler transformations: 
$L\to e^{-(f-\bar f)/2}\, L$. It is then convenient to define covariant derivatives 
$\nabla$ which include the $U(1)$ K\"ahler connection in addition to the standard 
metric-compatible Christoffel connection. On a scalar quantity of weight $p$, for instance,
one has $\nabla_i = \partial_i + p K_i$ and $\nabla_{\bar \imath} = \partial_{\bar \imath} - p K_{\bar \imath}$.
The covariant derivatives of $L$ are then found to be:  
\be
\nabla_{\bar \imath} L =0 \ ,\qquad
\nabla_{i} L =e^{K/2} \big(W_i + K_i W \big)\ .
\ee
From here one can see that $L$ is covariantly anti-holomorphic with its holomorphic 
covariant derivative being related to the order parameters of supersymmetry breaking.
Indeed the supersymmetry transformation of the fermions include the term 
$\delta_\epsilon \bar\chi^{\bar\imath} = -\sqrt{2}\,\bar\epsilon\, g^{\bar\imath j} N_j +\ldots\,$,
where the fermionic shifts $N_i$ are given by:
\be\label{chitrans}
N_i \equiv \nabla_i L\,.
\ee
For future reference let us also record the anti-holomorphic derivatives of $N_i$. These are simply
given by
\be
\nabla_{\bar \jmath}\, N_i = g_{i \bar \jmath} \, L \;\Rightarrow
\; \nabla_i \bar N^j = \delta_i^j \bar L\, .
\ee

Also note that since $\nabla_i$ involve both the Christoffel connection and the $U(1)$ K\"ahler connection, 
the commutator of two covariant derivatives acting on an object of non-zero $U(1)$ weight has an 
additional piece coming from the $U(1)$ curvature. For instance, on the fermionic shift one has: 
\be
\big[\nabla_i,\nabla_{\bar \jmath} \big] N_k = R_{i \bar \jmath k \bar s} \bar N^s - g_{i \bar \jmath} N_k \,,
\label{commid}
\ee
where $R_{i \bar \jmath k \bar s}$ is the Riemann tensor of the
K\"ahler manifold. (Our curvature conventions are summarized in Appendix~B.)

\subsection{Mass matrices}

Using the notation that we just introduced, the scalar potential $V$ takes the following simple form:
\be\label{Vdef}
V = \bar N^i N_i - 3 |L|^2 \,.
\ee
Its first derivative is then given by:
\be\label{stationary}
\nabla_i V = -2 N_i \bar L + \bar N^j \nabla_i N_j \,.
\ee 
Stationary points satisfy $\nabla_iV = 0$, and correspond to values of the scalar fields 
for which $\bar N^j\, \nabla_i N_j  = 2 N_i \bar L$.

Let us now compute the bosonic and fermionic mass matrices at a generic stationary point. 
The scalar masses are given by the second derivatives of $V$. These are easily computed 
and can be partly simplified by using the identity (\ref{commid}). One finds:
\be\begin{aligned}\label{ddV1}
\nabla_i \nabla_{\bar \jmath} V &= - 2 g_{i \bar \jmath} |L|^2\! + \nabla_i N_k \nabla_{\bar \jmath} \bar N^k \!
- R_{i \bar \jmath p \bar q} N^p \bar N^{\bar q} \! + g_{i \bar \jmath} \bar N^k N_k - N_i \bar N_{\bar \jmath} \,, \\[1mm]
\nabla_i \nabla_j V &= - \nabla_i N_j \bar L + \bar N^k \nabla_{(i} \nabla_{j)} N_k \,.
\end{aligned}
\ee
The two independent blocks for the mass matrix are then given by:
\be\label{mscalN1}
m_{0i \bar \jmath}^2 =  \nabla_i \nabla_{\bar \jmath} V \;,\qquad
m_{0i j}^2 =  \nabla_i \nabla_j V \,.
\ee

The fermionic mass matrix is also easy to compute. The mass terms for the physical fermions 
and the gravitino field can be read off from the following fermionic terms in the Lagrangian:
\be\begin{aligned}
{\cal L}_{\rm fm} =& - L \psi^\mu \sigma_{\mu \nu} \psi^\nu - \bar L \bar \psi^\mu \bar \sigma_{\mu \nu} \bar \psi^\nu
-  \tfrac  i{\sqrt{2}}\, N_i \chi^i \sigma_\mu \bar \psi^\mu 
+ \tfrac i{\sqrt{2}}\, \bar N_{\bar \jmath} \bar \chi^j \bar
\sigma_\mu \psi^\mu   \\
& - \tfrac 12 M_{ij} \chi^i \chi^j
- \tfrac 12 \bar M_{\bar \imath \bar \jmath}\chi^{\bar \imath}
\chi^{\bar \jmath} + \ldots\  ,
\end{aligned}
\ee
where
\be\label{Mdef}
M_{ij} \equiv \nabla_i N_j = \nabla_i \nabla_j L \,.
\ee
In the ground state the gravitino $\psi^\mu$ can be disentangled from the chiral fermions by the redefinition
\be
\tilde \psi^\mu = \psi^\mu + \tfrac i{3\sqrt{2}}\, L^{-1} N_{\bar \jmath} \, \sigma^\mu \chi^{\bar \jmath} \,,
\ee
where $L$ and $N_{\bar \jmath}$ are evaluated at the minimum of $V$.
This results in the following mass matrices for the physical fields: 
\be\begin{aligned}\label{physicalm}
m_{3/2} = L \,,\qquad 
m_{1/2ij} = M_{ij} - \tfrac 23 L^{-1} N_i N_j = \nabla_i N_j - \tfrac 23 L^{-1} N_i N_j \,,
\end{aligned}
\ee

The mass matrices (\ref{mscalN1}) and (\ref{physicalm}) obey a supersymmetric sum rule, which 
we record in Appendix~\ref{sumruleN1}.

\subsection{Goldstino and sGoldstinos}

As we already recalled, supersymmetry is spontaneously broken if in the vacuum $N_i \neq 0$. 
The associated Goldstino fermion is then given by the linear combination $\eta = N_i \chi^i$. 
This can be seen from the non-linear supersymmetry transformation of $\eta$ and/or from the fact 
that in a Minkowski vacuum the vector $N_i$ is a null vector of the physical mass matrix $m_{1/2ij}$. 
Indeed, from \eqref{Mdef} and the stationarity condition following from \eqref{stationary} it is easy to see that 
$M_{ij} \bar N^j = 2 \bar L\, N_i$. Using then \eqref{Vdef} and \eqref{physicalm} this implies
\be\label{meigen}
m_{1/2ij} \bar N^j = - \tfrac 23 L^{-1} V N_i \,,
\ee
with the right hand side being zero when $V=0$. The Goldstino field $\eta = N_i \chi^i$ has therefore 
a mass parameter which vanishes in Minkowski space and has a fixed value in units of the cosmological 
constant in AdS space:
\be
m_\eta =  - \tfrac 23 m_{3/2}^{-1} V \,.
\ee
Finally the complex sGoldstino, i.e.\ the scalar field describing the supersymmetric partners of
the Goldstino, is defined analogously by $\tilde \eta \equiv N_i \phi^i$.

\subsection{Stability of supersymmetric vacua}

Although in this paper we are interested in the stability of ground states with spontaneously broken supersymmetry, 
let us briefly present the proof that supersymmetric ground states are always stable. In this case one has
$N_i = 0$, which automatically implies the stationarity condition coming from \eqref{stationary} and a 
semi-negative definite cosmological constant $V = - 3 \,|L|^2$. Moreover, the scalar mass matrix simplifies as follows:
\bea
m^2_{0i \bar \jmath} = \nabla_i N_k \nabla_{\bar \jmath} \bar N^k - 2 g_{i \bar \jmath} |L|^2 \,, \qquad
m^2_{0i j} = - \bar L \nabla_i N_j  \,.
\eea
Looking  along an arbitrary direction $v^I=(v^i,v^{\bar \imath})$ in
field space with normalization $v^I v_I = 1$ (or $v^i v_i = 1/2$), one finds:
\bea
m^2_0 &=& m^2_{0 I \bar J} \,v^I v^{\bar J} = 2\,m^2_{0i \bar \jmath}\, v^i v^{\bar \jmath} + m^2_{0i j}\, v^i v^j 
+ m^2_{0\bar \imath \bar \jmath} \,v^{\bar \imath} v^{\bar \jmath} \nn \\
&=& \tfrac 12 \big(2 v^i \nabla_i N_k - v_k L \big) \big(2 v^{\bar \jmath} \nabla_{\bar \jmath} \bar N^k - v^k \bar L \big)
- \tfrac 92 \, v^i v_i \,|L|^2 \,.
\eea
In the last expression, the first term gives a semi-positive definite contribution so that $m_0^2$
satisfies the BF \cite{BF} bound\footnote{In $\textrm{AdS}_{d}$ the BF bound 
is given by $m^2 R^2 \ge -\tfrac14 (d-1)^2$, where $R$ is the AdS radius. For $d=4$ and $R^2
= -3 V^{-1}$ this leads to the bound \eqref{BFbound}.}
\be\label{BFbound}
m_0^2 \ge \tfrac  34\, V\,.
\ee
Notice that a minimal $m_0^2$ which saturates the bound can  only be achieved along the special complex 
directions $v_0^i$ for which the semi-positive terms are zero. These directions correspond to pseudo-eigenvectors 
of the matrix $M_{ij}$, in the sense that $M_{i}^{\;\bar \jmath} v_{0\,\bar \jmath}  = 2 L\, v_{0\,i}$. 

\subsection{Stability of non-supersymmetric vacua}

Let us now turn to the stability of non-supersymmetric vacua, that is, those for which
$N_i \neq 0$. This is largely discussed in refs.~\cite{GRS,CGGLPS} and here we only briefly
recall the results. However we do extend our previous analysis in that we also include 
non-supersymmetric AdS ground states.  

As was explained in detail in refs.~\cite{GRS,CGGLPS} the most stringent
constraints on the stability of the ground state come from the directions of the two sGoldstinos. 
Therefore we focus on the sGoldstino subspace defined by the complex direction $N_i$ and
consider the quantity
\bea
m_{\tilde \eta}^{2} \equiv \frac {m^2_{0i \bar \jmath} \bar N^i \hspace{-1pt} N^{\bar \jmath} \!}{\bar N^k N_k} \,.
\eea
With the help of \eqref{Vdef}, \eqref{stationary} and \eqref{ddV1} this can be rewritten as 
\bea\label{sGoldstinomass}
m_{\tilde \eta}^{2} &=& R_{\tilde \eta} \bar N^i N_i + 2 |L|^2 
= 3 \big(R_{\tilde \eta}+\tfrac 23\big) |m_{3/2}|^2 + R_{\tilde \eta} \, V \,,
\eea
where $R_{\tilde \eta}$ is the normalized holomorphic sectional
curvature along the sGoldstino direction, namely
\be
R_{\tilde \eta} = - \frac {R_{i \bar \jmath p \bar q} \bar N^i N^{\bar \jmath} \bar N^p N^{\bar q}}{(\bar N^k N_k)^{2}} \,.
\label{R}
\ee

The crucial observation is that $m_{\tilde \eta}^{2}$ represents an upper bound for the value of the 
smallest eigenvalue of the full mass matrix \cite{GRS,CGGLPS}.\footnote{In fact, the quantity $m_{\tilde \eta}^{2}$ arises as
half of the trace of the two-dimensional submatrix of the full
mass matrix along the two independent real directions
that can be formed out of the complex Goldstino direction.
It thus corresponds to the average of the two sGoldstino
square masses. The splitting of these two masses depends
explicitly on the superpotential and its derivatives, and is
therefore less interesting.}
 Therefore a necessary condition for 
stability is that the value of $m_{\tilde \eta}^{2}$ should be non-negative for dS or Minkowski vacua and 
should satisfy the BF bound \eqref{BFbound} for AdS vacua. It is convenient to 
phrase the discussion in terms of the following dimensionless parameter $\gamma$ defined as
\be
\gamma \equiv \frac {V}{3\,|m_{3/2}|^2} \,.
\label{gamma}
\ee
Minkowski/dS vacua correspond to $\gamma\in [0,+\infty)$ while AdS vacua have $\gamma \in [-1,0]$ 
since the cosmological constant is bounded to be larger than its critical supersymmetric value 
$V \ge - 3 |m_{3/2}|^2$.  Stability requires $m_{\tilde \eta}^2 \ge 0$ for dS vacua and 
$m_{\tilde \eta}^2 \ge \tfrac34\, V$ for AdS vacua which, using \eqref{sGoldstinomass} and \eqref{gamma}, 
can be viewed as the following bound for $R_{\tilde
\eta}$:\footnote{Note that we use here and in \eqref{R} a different sign 
convention for the Ricci-, scalar- and sectional curvatures of K\"ahler manifolds compared to 
refs.~\cite{GRS,CGGLPS}, although the Riemann tensor is defined in the same way. 
This is needed to consistently compare with the corresponding quantities for quaternionic-K\"ahler 
manifolds arising in next section. See Appendix~B for more details.}
\bea
R_{\tilde \eta} \ge 
\left\{\begin{array}{l}
\displaystyle{-\frac 23 \frac 1{1 + \gamma} \hspace{9pt} \quad \textrm{for}\quad \gamma \ge 0 \,,} \\[3mm]
\displaystyle{-\frac 23 \frac {1 - \tfrac98\, \gamma}{1 + \gamma} \quad \textrm{for}\quad \!\! -1\le \gamma \le 0 \,.}
\end{array}
\right.
\label{condR}
\eea

From this expression we see that the condition for finding metastable vacua with 
broken supersymmetry becomes more and more restrictive as the cosmological 
constant is increased: AdS vacua with minimal cosmological constant ($\gamma \to -1$) 
can always exist, as in such a case the condition simply reads $R_{\tilde \eta} > - \infty$. 
On the other hand, Minkowski vacua ($\gamma = 0$) can exist only if $R_{\tilde \eta} \ge - \tfrac 23$. 
Finally, dS vacua with large cosmological constant ($\gamma \to +\infty$) can exist only if
$R_{\tilde \eta} \ge 0$. The maximal freedom is therefore obtained in those models in which 
the sectional curvature $R_{\tilde \eta}$ either vanishes or turns out to be positive. 

Notice finally that in the limit in which gravity is decoupled by sending the Planck scale to infinity while
keeping the other scales fixed, the value of the quantity $m_{\tilde \eta}^2$ simplifies 
to the following expression:
\be
m_{\tilde \eta}^2 \simeq 3 R_{\tilde \eta} (1 + \gamma) |m_{3/2}|^2 \,.
\label{mrigid}
\ee
The standard limit of rigid supersymmetry can be obtained by further sending $m_{3/2}$
to zero. In that limit one finds $m_{\tilde \eta}^2 \simeq R_{\tilde \eta} V$.

Another interesting thing to note is the fact that the product of several K\"ahler-Hodge 
manifolds is again a K\"ahler-Hodge manifold. Thanks to this property, it is actually easy to 
construct models satisfying the necessary condition (\ref{condR}).
Indeed, starting with some manifolds ${\cal M}_i$ with sectional curvatures that are negative 
and bounded by some finite maximal values $R_i$, one can construct the product manifold 
${\cal M} = \times_i {\cal M}_i$ and find directions along which the sectional curvature is still 
negative but larger (i.e.\ closer to zero) than any of the individual $R_i$, the maximal possible 
value being $R_{\rm min} = (\sum_i R_i^{-1})^{-1}$. This means in particular that, by taking 
sufficiently many copies of any given K\"ahler manifold, one can always satisfy the condition~(\ref{condR}).

The fact that a K\"ahler manifold can factorize into several K\"ahler submanifolds also allows 
for situations in which the scalar fields spanning some of the submanifolds are stabilized in a 
supersymmetric way whereas the scalar fields spanning the rest of the submanifolds 
spontaneously break supersymmetry, provided that the superpotential also has some special properties. 
For the non-supersymmetric sector, one would get again a condition like (\ref{condR}), 
where $R_{\tilde \eta}$  now refers
to the relevant supersymmetry-breaking submanifold. 
For the supersymmetric sector, on the other hand, one should be careful as stability is not 
guaranteed in this case due to the fact that the cosmological constant is sourced by the other 
sector and departs from its critical supersymmetric value. 
As was shown in \cite{AHS1} for the particular case in which the two sectors interact only 
gravitationally, this cases cannot be viewed in general as a continuous limit of a supersymmetry 
breaking situation and therefore the stability of such vacua must then be studied separately.

\section{${\cal N}\!=2$ theories with hypermultiplets} \setcounter{equation}{0}

So far we have reviewed the stability of non-supersymmetric ground states
in ${\cal N}\!=1$ supergravity. Now we will move to the main topic of this paper 
and we will extend this analysis to the case of ${\cal N}\!=2$ supergravity coupled 
to an arbitrary number of hypermultiplets.

\subsection{Preliminaries}

Let us begin by reviewing the relevant aspects of the ${\cal N}\!=2$
theories and fix some conventions. For more details see, for example,
refs.~\cite{BW2,ABCDFFM,DF,ACDV}.\footnote{In the following we discuss 
gauged ${\cal N}\!=2$ supergravity in the standard electric frame following 
refs.~\cite{BW2,ABCDFFM,DF,ACDV}. In principle it is also possible to 
gauge with respect to the magnetic graviphoton (see, for example,
refs.~\cite{magnetic}). However, if only the graviphoton is present, the
symplectic rotation connecting the two cases is trivial and thus,
without loss of generality, we can confine our discussion to the
electric case.}
The gravitational multiplet contains the space-time metric
$g_{\mu\nu}$, a pair of gravitini $\psi_\mu^A,\, A=1,2$ and an
Abelian graviphoton $A_\mu$. This multiplet can be coupled to $n$
hypermultiplets $H^i,\, i=1,\ldots,n$ 
which contain $4n$ scalar fields $q^u,\, u=1,\ldots,4n$ and $2n$ 
fermions $\xi^{\alpha }\,, \alpha=1,\ldots,2n$.  The scalar fields $q^u$ span 
a quaternionic-K\"ahler manifold of dimension $4n$  with 
holonomy group $Sp(2n)\times SU(2)$.

On a quaternionic-K\"ahler manifold there exists a triplet of almost
complex structures $J^x,\,x=1,2,3$ which satisfy an $SU(2)$ algebra.
Associated with them is a triplet of Hyperk\"ahler two-forms $\Omega^x$
which consequently obey
\be
\Omega^x_{u w} \Omega^{yw}_{\;\;\;\; v} = - h_{uv} \delta^{xy} - \epsilon^{xyz} \Omega^z_{uv} \,,
\ee
where $h_{u v}$ is the quaternionic metric. Furthermore, the $\Omega^x$
are identified with the field strength of the $SU(2)$ part of the holonomy group
and as a consequence they are covariantly constant with respect to the
$SU(2)$ connection: $\nabla_w\Omega^x_{uv}=0$.\footnote{Strictly speaking 
$\Omega^x$ only needs to be proportional to the  Hyperk\"ahler two-forms but in order to
simplify the notation we have chosen the proportionality factor to be equal to $-1$, as is usually 
done in the literature (corresponding to $\lambda=-1$ in \cite{ABCDFFM,DF} and $\nu = -2$ in \cite{ACDV}).}
Here and in the following, $\nabla_u $ denotes a covariant derivative involving also the $SU(2)$ 
connection.

Our conventions are as follows. The $SU(2)$ doublet indices $A,B$ are raised and lowered in 
the usual way with the antisymmetric tensors $\epsilon_{AB}$ and $\epsilon^{AB}$ and the matrices 
$\sigma^{x\;\;B}_{\;A}$ denote the usual antisymmetric Pauli matrices. The matrices $\sigma^x_{AB}$ 
and $\sigma^{xAB}$ are then symmetric and satisfy $(\sigma^x_{AB})^* = - \sigma^{xAB}$. They relate 
$SU(2)$ triplets to the symmetric product of two $SU(2)$ doublets, and can be used to alternatively 
describe any triplet as a bi-doublet through the definition $\xi^{AB} \equiv i \xi^x \sigma^{xAB}$. 
Moreover, they satisfy the identity:
\be
\sigma^x_{AB} \sigma^x_{CD} = 2 \epsilon_{A(C} \epsilon_{BD)} \,.
\label{magic}
\ee
For the $Sp(2n)$ group, we denote by $\alpha,\beta = 1,\dots, 2n$ the
$2n$-plets indices. 
These are raised and lowered with the antisymmetric symplectic tensors $C_{\alpha \beta}$ 
and $C^{\alpha \beta}$.

It is convenient to define a vielbein $U_u^{\alpha A}$ for the quaternionic metric 
by the relation $h_{u v} = U_u^{\alpha A} U_v^{\beta B} \epsilon_{AB} C_{\alpha \beta}$.
The inverse vielbein $U^u_{\alpha A}$ then satisfies $U^u_{\alpha A} U_v^{\alpha A} = \delta^u_v$ 
and $U_u^{\alpha A} U^u_{\beta B} = \delta^\alpha_\beta \delta^A_B$. These actually
satisfy the stronger relations $h_{u v} = \epsilon_{AB}  U_u^{\alpha A} U_{v\alpha}^{B}$ and 
$\Omega^x_{u v} = - i \sigma^x_{AB}  U_u^{\alpha A} U_{v\alpha}^{B}$, or
$U^u_{\alpha A} U^v_{\beta B} h_{u v} = \epsilon_{AB} C_{\alpha \beta}$ and 
$U^u_{\alpha A} U^v_{\beta B} \Omega^x_{u v} = - i \sigma^x_{AB} C_{\alpha \beta}$,
which are conveniently summarized in the identity:
\be
U_u^{\alpha A} U^B_{\alpha v} = \tfrac 12 h_{uv} \epsilon^{AB}\! - \tfrac i2 \Omega_{uv}^x \sigma^{xAB}
\label{UUident} \,.
\ee

The curvature consists of an $SU(2)$ part and an $Sp(2n)$ part with the corresponding curvature 
forms given by:
\be
R^{AB}_{uv} = - i\, \Omega^x_{uv} \sigma^{xAB} \,,\qquad
R^{\alpha \beta}_{uv} = \epsilon_{AB} U_{[u}^{\gamma A} U_{v]}^{\delta B} 
\big(\!-\! 2\, \delta^\alpha_{(\gamma} \delta^\beta_{\delta)} + \Sigma^{\alpha \beta}_{\;\;\;\;\gamma\delta} \big)\,.
\ee
The tensor $\Sigma_{\alpha \beta \gamma \delta}$ must be completely symmetric but is otherwise arbitrary,
and represents the only freedom in the curvature. The full Riemann tensor with two `flat' index-pairs is 
given by $R^{\alpha A \beta B}_{\;\;\;\;\;\;\;\;\;\; u v} = R^{AB}_{uv} C^{\alpha \beta} + R^{\alpha \beta}_{uv} \epsilon^{AB}$.
Using eq.~(\ref{magic}) the curvature with only flat indices  is found to be 
\bea
R_{\alpha A \beta B \gamma C \delta D} 
&=& 2\,\epsilon_{A(C} \epsilon_{BD)} C_{\alpha \beta} C_{\gamma \delta}
+ \epsilon_{AB} \epsilon_{CD} \big(\!-\! 2\, C_{\alpha(\gamma} C_{\beta \delta)} + \Sigma_{\alpha \beta \gamma\delta} \big) \,.
\eea
Its version with only curved indices is instead given by:
\be
R_{u v r s} = - h_{u [r} h_{vs]} - \Omega^x_{uv} \Omega^x_{rs} - \Omega^x_{u[r} \Omega^x_{vs]}
+ \Sigma_{u v r s} \,,
\label{Riemann}
\ee
where:
\be
\Sigma_{u v r s} = \epsilon_{AB} \epsilon_{CD} U_u^{\alpha A} U_v^{\beta B} U_r^{\gamma C} U_s^{\delta D} 
\Sigma_{\alpha \beta \gamma \delta} \,.
\ee
The tensor $\Sigma_{u v r s}$ behaves like a Weyl component of the Riemann tensor, in the sense that 
any contraction with the metric vanishes. This implies that the Ricci tensor is completely universal and 
that quaternionic-K\"ahler manifolds are Einstein manifolds with
\be
R_{u v} = - 2 (n + 2) h_{u v} \,, \qquad R = - 8 n (n + 2) \,.
\label{Ricci}
\ee

So far we have discussed the ungauged ${\cal N}\!=2$ theory. Let us now turn to the
situation in which an isometry of $h_{uv}$ is gauged with the graviphoton 
$A_\mu$. In this case the scalars are charged under the isometry group and transform
as $\delta q^u  =\Lambda\, k^u(q)$, where $\Lambda$ is the space-time dependent 
gauge parameter while $k^u(q)$ is the Killing vector, which satisfies the Killing equation
\be\label{Killing}
\nabla_{(u} k_{v)} = 0\, .
\ee 
In the Lagrangian all the space-time derivatives acting on scalar fields are then replaced by covariant
derivatives, of the form $D_\mu q^u \equiv \partial_\mu q^u + k^u A_\mu$. 

On a quaternionic-K\"ahler manifold, any Killing vector $k^u$ can be expressed in
terms of a triplet of Killing potentials $P^x$, defined by
\be
\nabla_u P^x = 2\,\Omega^x_{uv} k^v \,.
\label{relPk}
\ee
Actually one can also relate $k^u$ and $P^x$ as:
\be
k_u = - \tfrac 16\, \Omega^x_{u v} \nabla^v P^x \,,\qquad
P^x = \tfrac 1{2 n}\, \Omega^x_{u v} \nabla^u k^v \,.
\ee
One also finds the following relations for the second derivatives of these quantities:
\be\begin{aligned}
 \big[\nabla_u, \nabla_v\big] P^x &= 2\,\epsilon^{xyz} \Omega^y_{u v} P^z \,, \label{comm} \\
\big[\nabla_u, \nabla_v\big] k_w &= R_{uvws} k^s \,, \\
\nabla_u \nabla_v k_w &= - R_{vwus} k^s \,.
\end{aligned}\ee
Moreover, $P^x$ and $k_u$ satisfy the harmonic equations
\be
\nabla^w \nabla_w P^x = 4 n P^x \,, \qquad
\nabla^w \nabla_w k_u = 2 (n + 2)\, k_u \,.
\ee

Finally, the derivatives of the Killing potentials $P^x$ turn out to be related to the order parameters
of supersymmetry breaking. Indeed, the supersymmetry transformation of the hyperini has the 
form $\delta \xi_\alpha = N_\alpha^A\epsilon_A +\ldots\ $, where the fermionic shifts $N_\alpha^A$
are given by:
\be
N^A_\alpha = 2 U_{u \alpha}^A k^u = \tfrac 13 U_{\alpha B}^u \nabla_u P^{AB} \,.
\ee

\subsection{Mass matrices}

The scalar potential can be expressed in terms of the Killing vector and the Killing potentials, and 
takes the following simple form:
\bea
V &=& N_A^\alpha N^A_\alpha - 3\, P^x P^x = 4\, k^w k_w - 3\, P^x P^x \,.
\eea
Its first derivatives are given by
\bea
\nabla_u V &=& 8\, k^w \nabla_u k_w - 6\, P^x \nabla_u P^x \,,
\label{dV}
\eea
and stationary points where $\nabla_u V = 0$ are thus characterized by the condition
$k^w \nabla_u k_w = \tfrac 34 P^x \nabla_u P^x$.

The scalar mass matrix at a stationary point of the potential is related to the second 
derivatives of the potential. These are found to be:
\bea
\nabla_u \nabla_v V = 8\, \nabla_u k^w \nabla_v k_w - 8\, R_{u s v t} k^s k^t \!
- 6\, \nabla_u P^x \nabla_v P^x \! - 6\, P^x \nabla_{(u} \nabla_{v)} P^x \,.\
\label{ddV}
\eea
In the conventions we are following, the kinetic term of the scalar fields has the non-canonical 
form ${\cal L}_{\rm kin} = - h_{uv} D_\mu q^u D^\mu q^v$. The properly normalized mass matrix
for the scalars is thus given by: 
\be\label{mscalar2}
m^2_{0uv} = \tfrac 12 \nabla_u \nabla_v V \,.
\ee

The square mass of the graviphoton is induced by the connection terms in the covariant derivatives 
of the scalars kinetic term. Taking into account that with the conventions we are following the kinetic 
term for the graviphoton has the non-canonical form ${\cal L}_{\rm kin}=-\tfrac18 F_{\mu\nu}F^{\mu\nu}$, 
one deduces that: 
\be
m_1^2 = 4 k^u k_u \,.
\ee

The mass terms for the hyperini and the gravitini can be read off from
the fermionic part of the ${\cal N}\!=2$ Lagrangian
\bea
{\cal L}_{\rm fm} &=& P_{AB} \bar \psi^A_\mu \gamma^{\mu \nu} \psi_\nu^B 
+ \bar P^{AB} \bar \psi_{A \mu} \gamma^{\mu \nu} \psi_{\nu B} 
+ 2 i N_\alpha^A \bar \xi^\alpha \gamma_\mu \psi^\mu_A 
+ 2 i  N^\alpha_A \bar \xi_\alpha \gamma_\mu \psi^{\mu A} \nn \\[1mm]
&\;& +\, M_{\alpha \beta} \bar \xi^\alpha \xi^\beta + \bar M^{\alpha
\beta} \bar \xi_\alpha \xi_\beta +\ldots \,,
\eea
where
\be\begin{aligned}
M_{\alpha \beta} &= - U^u_{\alpha A} U^v_{\beta B} \epsilon^{AB} \nabla_{[u} k_{v]} 
= - \tfrac 16 \, U^u_{\alpha A} U^v_{\beta B} \nabla_u \nabla_v P^{AB} \,.
\end{aligned}\ee
In order to disentangle the gravitino from the Goldstino, 
one redefines
\be
\tilde \psi^{\mu A} = \psi^{\mu A} + \tfrac {i}{3} P^{-1AB}  N^\beta_B \gamma^\mu \xi_\beta \,,
\ee
which results in the following mass matrices for the physical fermions and the two
gravitini\footnote{Notice that $P_{AC} \bar P^{CB} = P^x P^x \, \delta_A^B$.
It follows that $P^{-1AB} = (P^x P^x)^{-1} \bar P^{AB}$ and similarly 
$\bar P^{-1}_{AB} = (P^x P^x)^{-1} P_{AB}$. }
\bea
m_{1/2 \alpha \beta} &=& M_{\alpha \beta} - \tfrac 43 \bar P^{-1}_{AB} \,N_\alpha^A N_\beta^B
= - U^u_{\alpha A} U^v_{\beta B} \Big(\epsilon^{A B} \nabla_{[u} k_{v]} 
+ \tfrac {16}{3} P^{AB} |m_{3/2}|^{-2} k_u k_v \Big) \,, \nn \\
m_{3/2AB} &=& P_{AB} \,. \label{mphys2}
\eea
Thus, the gravitino mass scale is simply given by:
\be
|m_{3/2}| = \sqrt{P^x P^x} \,.
\ee

Comparing with the formulation of ${\cal N}\!=1$ theories described in Section 2, we can now identify the generalization 
of each ingredient to the ${\cal N}\!=2$ case. We see that $P^x$ is the generalization of $L$ while $N_\alpha^A$ is 
instead the generalization of $N_i$.

\subsection{Goldstinos and sGoldstinos}

Supersymmetry is spontaneously broken whenever $N^A_\alpha \neq 0$ on the vacuum. 
The corresponding two Goldstino fermions are then given by
$\eta^A = N^A_\alpha \xi^\alpha$. Using the stationarity condition following from \eqref{dV} 
and the properties of the vielbein one can show that
\be
M_{\alpha \beta} N^\beta_A = 2\, P_{AB} N^B_\alpha \,. 
\label{rel}
\ee
Using  (\ref{mphys2}) together with the relation $ N_A^\alpha N^B_\alpha
= 2 \,k^w k_w \, \delta_A^B$, eq.~\eqref{rel} implies
\be
m_{1/2\alpha \beta}\,  N^\beta_A = - \tfrac 23\, V \bar P^{-1}_{AB} N^B_\alpha \,.
\ee
Thus we see again that the normalized mass matrix for the two Goldstinos vanishes identically in 
Minkowski space and has a fixed form in units of the cosmological constant in AdS space:
\be
m_{\eta AB} = - \tfrac 23 m_{3/2 AB}^{-1} V\,.
\ee

The two independent Goldstino fermions $\eta^A = N^A_\alpha \xi^\alpha$, which transform as a 
doublet under $SU(2)$, have four real sGoldstino partners given by $\tilde \eta^{AB} = N^{AB}_u q^u$. 
The quantity $N^{AB}_u$ transforms as the tensor product of two $SU(2)$ doublets, and can be computed 
by acting with the inverse vielbein $U^{\alpha A}_u$ on  $N_\alpha^B$. This is a result of the fact that  
$U^{\alpha A}_u$ locally maps the tangent space where the fermions are defined to the coordinates 
of the manifold associated with the scalar fields. More precisely, one finds:
\be
N_u^{AB} = U_u^{\alpha A} N_\alpha^B = N_u\, \epsilon^{AB} + i N_u^x\,\sigma^{xAB} \,, 
\ee
where in the second equation we used the identity (\ref{UUident}) to decompose
$N_u^{AB}$  into a singlet $N_u$ plus a triplet  $N_u^x$ with
\bea
N_u = k_u \;,\qquad
N_u^x = - \Omega^{xv}_u k_v = - \tfrac 12 \nabla_u P^x \,.
\eea
The four-dimensional space of sGoldstino directions can thus be
parameterized by $(N_u,N_u^x)$.\footnote{Note that $N_u^x$ conjugates
$N_u$ with respect to each of the three almost complex structures
$\Omega^{xv}_u$.} 
These vectors form an orthonormal basis, in the sense that:
\bea
N^u N_u = k^u k_u \;,\qquad
N^{xu} N^y_u = k^u k_u \, \delta^{xy} \;, \qquad
N^u N^x_u = 0 \,.
\eea 
It is then convenient to use the fields $\tilde \eta = N_u q^u$ and $\tilde \eta^x = N_u^x q^u$ to 
parameterize the four independent sGoldstinos.

\subsection{Stability of supersymmetric vacua}

Let us consider first the case of supersymmetric vacua. Unbroken supersymmetry implies 
\be
k_u = 0 \;\Rightarrow\; N_u = N_u^x = 0 \,.
\ee
As usual, any point in the scalar field space where these conditions are fulfilled is automatically 
a stationary point of the potential, as can be seen from eq.~(\ref{dV}). At such points 
the cosmological constant is negative and given by  $V=-3 P^x P^x$. Moreover, the 
mass matrix (\ref{mscalar2}) simplifies and can be rewritten as
\be\begin{aligned}
m^2_{0uv} &= 4\, \nabla_u k^w \nabla_v k_w - 3\, P^x \nabla_{(u} \nabla_{v)} P^x  \\
&= 4 \Big(\nabla_u k^w - \tfrac 34 P^x \Omega^{xw}_{u} \Big) \Big(\nabla_v k_w - \tfrac 34 P^y \Omega^y_{v w} \Big) 
- \tfrac 94\, h_{u v} P^x P^x \,.
\label{V2susy}
\end{aligned}\ee
In the last expression, the first term is semi-positive definite, so the value of the mass matrix along any normalized 
direction $v^u$, with $v^u v_u=1$, satisfies the BF bound \eqref{BFbound} which guarantees
stability: $m^2_{0} = m^2_{0uv} v^u v^v \ge \frac 34\, V$.
Note that the directions $v_0^u$ in field space for which this bound is saturated satisfy an 
equation of the form $(\nabla_u k_v) v_0^v = \tfrac34\, P^x \Omega^{x}_{uv} v_0^v$.

\subsection{Stability of non-supersymmetric vacua}

Let us now study the conditions under which metastable
non-supersymmetric vacua can exist. 
Spontaneously broken supersymmetry implies
\be
k_u \neq 0 \;\Rightarrow\; N_u, N_u^x \neq 0 \,.
\ee
One can then study the mass matrix in the four-dimensional subspace of sGoldstino directions 
spanned by the four vectors $N_u=k_u$ and $N_u^x= -\Omega_u^{xv} k_v$. Gauge invariance of the potential 
implies however that at any stationary point the vector $N_u$ is a flat direction of the potential, 
corresponding to the would-be Goldstone boson that is eaten by the graviphoton. Let us then 
study the mass matrix in the three-dimensional subspace defined by the vectors $N_u^x$ given by
\bea
m_{\tilde \eta}^{2xy} = \frac {m^2_{0uv} N^{ux} N^{v y}}{N^w N_w} \ .
\eea
This expression for $m_{\tilde \eta}^{2xy}$ can be simplified using equations (\ref{ddV}) and (\ref{mscalar2}) 
and the stationarity condition coming from \eqref{dV}.\footnote{The main intermediated step needed is the relation 
$\nabla^wk_u\nabla_wP^x = 3 P^x k_u + \tfrac12 \epsilon^{xyz}P^y\nabla_uP^z$. This can 
be derived by taking a derivative of the identity $k^w\nabla_wP^x = 0$ and using then the stationarity 
condition and the first relation in (\ref{comm}).} One then finds, after a straightforward
computation, the following simple expression
\be\label{scalarmm2}
m_{\tilde \eta}^{2xy} = - 4 \big(R_{\tilde \eta}^{xy} + 3\, \delta^{xy} \big) k^w k_w + 4 \big(\delta^{xy} - \pi^{xy} \big) P^z P^z \,,
\ee
where 
\be\label{projector}
\pi^{xy} = \frac {P^x P^y}{P^z P^z} 
\ee
is the projector along the direction defined by $P^x$ and 
$R_{\tilde \eta}^{xy}$ is given by
\be
R_{\tilde \eta}^{xy} = \frac {R_{usvt} N^{ux} N^s N^{vy} N^t}{(N^w N_w)^2} \,.
\label{Rxy}
\ee
This quantity is something like a tri-holomorphic sectional curvature for the quaternionic
directions $N_u^{AB}$, in the sense that its diagonal elements correspond to the three independent
holomorphic sectional curvatures that one can build out of $N_u$ and one of its conjugates $N_u^x = J^x_{uv} N^v$.
Using the expression (\ref{Riemann}) for the Riemann tensor, one can evaluate  $R_{\tilde \eta}^{xy}$ more explicitly,
and express it in terms of the tensor $\Sigma_{\alpha \beta \gamma \delta}$. One actually finds:
\be
R_{\tilde \eta}^{xy} = - 2\, \delta^{xy} - \frac {\Sigma_{\alpha \beta \gamma \delta} 
N^{\alpha A} N^{\beta B} N^{\gamma C} N^{\delta D}}
{( N^{\epsilon E} N_{\epsilon E})^2 }\ \sigma^x_{AB} \sigma^y_{CD} 
\label{RSigmaxy} \,.
\ee

Metastability of the vacuum requires that the eigenvalues of the three-dimen\-sional matrix $m_{\tilde \eta}^{2xy}$ 
given in \eqref{scalarmm2} should be either positive or above the BF bound, depending on the sign of the cosmological constant.
This condition depends on the tensor $\Sigma_{\alpha \beta \gamma \delta}$ in a non-trivial way, and can be 
understood as a constraint on it. More precisely, it restricts the values that the curvature is allowed to take in the 
subspace of sGoldstino directions. As in the previous section, to analyze the implications of the metastability 
constraints it is convenient to parametrize the value of the cosmological constant through the dimensionless 
parameter $\gamma = V/(3|m_{3/2}|^2)$.

To study the matrix $m_{\tilde \eta}^{2xy}$, it is convenient to switch to a basis of eigenvectors of the projector 
$\pi^{xy}$, which we shall denote by $v_i^x$, $i=1,2,3$, for the eigenvalues $\lambda_i = (1,0,0)$
(so that $v_1^x$ is the direction defined by $P^x$ and $v_{2,3}^x$ span the subspace orthogonal to it).
These vectors can be chosen in such a way as to form an orthonormal and complete basis of the three-dimensional 
space under consideration, with:
\be
\label{defvi}
\begin{aligned}
& \pi^{xy} v_i^y = \lambda_i v_i^x \quad \mbox{(no sum on $i$)} \quad \mbox{and}\;\; \lambda_i = (1,0,0) \,, \\
& v_i^x v_j^x = \delta_{ij} \,,\;\; v_i^x v_i^y = \delta^{xy} \,.
\end{aligned}
\ee
In this new basis, the matrix $m_{\tilde \eta ij}^2 \equiv m^{2xy}_{\tilde \eta} v_i^x v_j^y$ is still not diagonal. 
But each of its diagonal elements must 
nevertheless necessarily satisfy the metastability bound on their own. These three elements define indeed the 
values of the square mass along the three special orthogonal directions $v_i^x N_u^x$, which we shall denote by:
\be
m_{\tilde \eta i}^2 \equiv m_{\tilde \eta}^{2xy} v_i^x v_i^y \quad \mbox{(no sum on $i$)} \,.
\ee
Using \eqref{scalarmm2} and \eqref{defvi} one computes
\be
\label{miR}
m_{\tilde \eta i}^2 
= - 3 \Big(R_{\tilde \eta i} + \tfrac 53 + \tfrac 43 \lambda_i \Big) |m_{3/2}|^2 - \Big(R_{\tilde \eta i} + 3\Big) V \,,
\ee
in terms of the holomorphic sectional curvatures defined by the rotated complex structures 
$J_{iuv} = J^x_{uv} v_i^x$, which are given by:
\bea
R_{\tilde \eta i} \equiv R_{\tilde \eta}^{xy} v_i^x v_i^y \quad \mbox{(no sum on $i$)} \,.
\label{Ri}
\eea
The metastability condition ($m_0^2 \ge 0$ if $V \ge 0$ and $m_0^2 \ge \tfrac 34 V$ if $V < 0$) applied to 
$m_{\tilde \eta i}^2$ then implies
\bea
R_{\tilde \eta i} \le 
\left\{\begin{array}{l}
\displaystyle{\!- \frac {5 \!+\! 4 \lambda_i}3\, \frac {1 \!+\! \tfrac 9{5 + 4 \lambda_i}\, \gamma}{1 + \gamma} \hspace{6pt} \,,\;\; \gamma \ge 0 \,,} \\[3mm]
\displaystyle{\!- \frac {5 \!+\! 4 \lambda_i}3\, \frac {1 \!+\! \tfrac {45}{4(5 + 4 \lambda_i)}\, \gamma}{1 + \gamma} \,,\;\; \gamma \le 0 \,.}
\end{array}
\right.
\label{condRi}
\eea

Summarizing, we see that in ${\cal N}\!=2$ theories we get three conditions, all similar 
to the one of ${\cal N}\!=1$. They are associated with three of the partners of 
the two independent Goldstinos. Note however that the coefficients in the quantities 
$m_{\tilde \eta i}^2$ differ from the coefficients in the ${\cal N}\!=1$ quantity 
$m_{\tilde \eta}^2$ given in \eqref{sGoldstinomass}. This is reasonable, since the geometry is 
quaternionic-K\"ahler for ${\cal N}\!=2$ and K\"ahler-Hodge for ${\cal N}\!=1$, and these 
two kinds of geometries are unrelated.\footnote{A notable exception to this general
fact is given by the family of coset manifolds $SU(2,n)/(U(1) \times SU(2) \times SU(n))$, 
which turn out to be both K\"ahler-Hodge and quaternionic-K\"ahler.} 
Furthermore, note that the sectional curvatures enter
\eqref{sGoldstinomass} and \eqref{miR} with a different sign, once compatible 
conventions for real and complex manifolds are used (see Appendix B). This results 
in opposite inequality signs in the metastability constraints on the
sectional curvature given in \eqref{condR} and \eqref{condRi}.

Before we proceed let us inspect the limit where gravity is decoupled by 
sending the Planck scale to infinity. In this limit, the ${\cal N}\!=2$ 
geometry becomes Hyperk\"ahler while the ${\cal N}\!=1$ geometry 
becomes K\"ahler. The two geometries are then related, in the sense 
that the former is just a subclass of the latter. As a result, 
${\cal N}\!=2$ theories reduce to a special case of ${\cal N}\!=1$ 
theories, and the metastability conditions arising in the two cases 
can be directly compared. In this rigid limit, however, in which the 
graviton, the gravitino and the graviphoton are decoupled, the scalar 
potential of ${\cal N}\!=2$ theories with only hypermultiplets becomes 
trivial. This corresponds to the fact that from the ${\cal N}\!=1$ perspective 
the superpotential vanishes. 
As a result also the sGoldstino masses go to zero, independently of 
the curvature of the Hyperk\"ahler manifold and we have
\be
m_{\tilde \eta i}^2 \simeq 0 \,.
\ee
This means that in this limit the ${\cal N}\!=2$ conditions can never really 
be satisfied, since the potential identically vanishes and thus the scalar fields 
cannot be stabilized.  
The ${\cal N}\!=1$ conditions implied by \eqref{mrigid}, on the other hand,
can be satisfied for models with suitable geometry, but when the 
superpotential is sent to zero the scalar masses flow to zero also 
in this case.

Up to now we have not used the fact  that quaternionic-K\"ahler
manifolds have a constrained  
curvature tensor with a sectional curvature given in (\ref{RSigmaxy}).
Similarly, the 
$R_{\tilde \eta i}$ that appear in \eqref{miR} take the form:
\bea
\label{RSigmai}
R_{\tilde \eta i} = -2 + \Delta_i(\Sigma) \,,
\eea
where
\be
\Delta_i(\Sigma)\ \equiv\ \frac {\Sigma_{\alpha \beta \gamma \delta}
N^{\alpha A} N^{\beta B}  N^{\gamma C} N^{\delta D}}{( N^{\epsilon E} N_{\epsilon E})^2}\
v_i^x \sigma^x_{AB} v_i^y\sigma^y_{CD} \quad \mbox{(no sum on $i$)} \ .
\ee
So the metastability conditions constrain the allowed values for the
quantities  
$\Delta_i(\Sigma)$, for a given value of the parameter $\gamma$.

As a first remark, note that for those particular quaternionic-K\"ahler manifolds for which 
the tensor $\Sigma_{\alpha \beta \gamma \delta}$ vanishes, the situation simplifies 
substantially.\footnote{This is for instance the case for the family of 
coset manifolds $Sp(2,2n)/(Sp(2) \times Sp(2n))$.}
Indeed, in that case one simply has $R_{\tilde \eta i} = - 2$, and thus 
$m_{\tilde \eta i}^2 = (1- 4 \lambda_i) |m_{3/2}|^2 - V$, that is 
$m_{\tilde \eta 1}^2 = -\, V\! -  3 \, |m_{3/2}|^2$ in the direction parallel to $P^x$ and 
$m_{\tilde \eta 2,3}^2 = -\, V\!+ |m_{3/2}|^2$ along the two directions orthogonal to $P^x$. 
These satisfy the stability bound only if $\gamma \in [-1,-\tfrac47]$, and 
thus Minkowski/dS vacua are excluded. 

Even for more general quaternionic-K\"ahler manifolds with $\Sigma\neq0$, we can actually obtain a 
stronger constraint from \eqref{miR}. Notice in this respect that the three square masses \eqref{miR} 
transform as a triplet under $SU(2)$ $R$-symmetry transformations, reflecting the fact that they 
are associated with the triplet of almost complex structures existing on quaternionic-K\"ahler manifolds. 
One may then try to look for an $SU(2)$ singlet projection and check whether it leads to any useful information. 
From the point of view of the original mass matrix  $m_{\tilde \eta}^{2xy}$, the only object that could lead to such 
a thing is the trace. More precisely, one can consider the average of the diagonal elements, which by the 
completeness relation in \eqref{defvi} also corresponds to the average of the three masses $m_{\tilde \eta i}^2$ 
computed above:
\be
m_{\tilde \eta}^{2} \equiv  \tfrac 13 \delta^{xy} m_{\tilde \eta}^{2x y} 
= \tfrac 13 \, \mbox{\small $\sum_i$} m_{\tilde \eta i}^2 \,.
\ee
Using \eqref{scalarmm2} one arrives at
\be
\label{mR}
m_{\tilde \eta}^{2} 
= - 3 \Big(R_{\tilde \eta} + \tfrac {19}9 \Big) |m_{3/2}|^2 - \Big(R_{\tilde \eta} + 3\Big) V \,, \\
\ee
where $R_{\tilde \eta}$ is the averaged sectional curvature 
\bea
R_{\tilde \eta} \equiv \tfrac 13\, \delta^{xy} R^{xy} = \tfrac 13 \, \mbox{\small $\sum_i$} R_{\tilde \eta i} \,.
\label{Rav}
\eea
Note now that $m_{\tilde \eta}^2$ also gives an upper bound on the smallest mass eigenvalue,
as a consequence of the fact that each $m_{\tilde \eta i}^2$ gives itself a lower bound.\footnote{Indeed,
$m_{\tilde \eta}^2$ is the averaged trace of the matrix, and gives thus the average of the eigenvalues. 
Each $m_{\tilde \eta i}^2$ is instead just the projection of the matrix along a specific direction, and 
is thus a combination of the eigenvalues with coefficients whose square sum up to $1$. 
In both cases, the resulting value is clearly an upper bound to the smallest 
eigenvalue of $m^{2xy}_{\tilde \eta}$, and thus also of the full mass matrix $m_{0uv}^2$.} The 
metastability condition applied to $m_{\tilde \eta}^2$ then implies
\bea
R_{\tilde \eta} \le 
\left\{\begin{array}{l}
\displaystyle{\!- \frac {19}9\, \frac {1 \!+\! \tfrac{27}{19}\, \gamma}{1 + \gamma} \hspace{6pt} \,,\;\; \gamma \ge 0 \,,} \\[3mm]
\displaystyle{\!- \frac {19}9\, \frac {1 \!+\! \tfrac{135}{76}\, \gamma}{1 + \gamma} \,,\;\; \gamma \le 0 \,.}
\end{array}
\right.
\eea
The crucial observation that one can make at this point is that the
averaged sectional curvature  
$R_{\tilde \eta}$ actually  
is independent of the tensor $\Sigma_{\alpha \beta \gamma \delta}$
and thus takes a universal value 
common to all the possible quaternionic-K\"ahler manifolds. 
Indeed, using the property \eqref{magic} in 
\eqref{RSigmai} and \eqref{Rav},  one finds:\footnote{This follows from 
the fact that the contraction $N^{\alpha A} N^{\beta B}\epsilon_{AB}$ is antisymmetric in $\alpha,\beta$ whereas 
the tensor $\Sigma_{\alpha \beta \gamma \delta}$ is completely symmetric in all indices.}
\be
\label{RSigma}
R_{\tilde \eta} = - 2 - \frac 23 \frac {\Sigma_{\alpha \beta \gamma \delta}
N^{\alpha A} N^{\beta B}  N^{\gamma C} N^{\delta D}}{( N^{\epsilon E} N_{\epsilon E})^2}\
\epsilon_{AB}  \epsilon_{CD}  = -2 \,.
\ee
Inserting \eqref{RSigma} into \eqref{mR} one then finds the simple expression
\be
m_{\tilde \eta}^2 = - \tfrac 13 \big(1 + 9\, \gamma \big) |m_{3/2}|^2 \,.
\ee
This satisfies the metastability bound only for 
\be\label{finalbound}
\gamma \in \big[\!-\!1,- \tfrac{4}{63}\big] \,.
\ee
Notice that this restriction implies in paticular that dS vacua are always
excluded.\footnote{This result was already know to hold for the particular subclasses of 
quaternionic-K\"ahler manifolds for which $n=1$ as well as those with $n> 1$ and 
$\Sigma_{\alpha \beta \gamma \delta} = 0$ \cite{DV}.}
This is unavoidable and holds true for any scalar geometry.\footnote{
There is an apparent counter-example of this result in ref.~\cite{DSTV}, where 
a metastable dS vacuum was found in the universal hypermultiplet geometry with 
instanton corrections taken into account.
However the approximation used does not keep the metric quaternionic and
we suspect that the dS vacuum is destabilized once the higher instanton
corrections required to make the metric quaternionic are included.
We understand that preliminary investigations point in this direction and
we thank F. Saueressig for discussions on this issue.}
AdS vacua, on the other hand, are allowed if they satisfy \eqref{finalbound}.
This represents the main result of our investigation.

Notice finally that the product of several quaternionic-K\"ahler manifolds is no longer
a quaternionic-K\"ahler manifold. This is a consequence of the particular form that the Riemann
curvature tensor must take. More precisely, the Ricci- and scalar curvatures are completely 
fixed by the dimensionality of the space (c.f.\ \eqref{Ricci}), and this relation is destroyed when taking the 
product of two of such manifolds. Thus, there is no easy way of diluting the curvature just by taking 
products of manifolds and thus the bound is always unavoidably violated.

\section{Conclusions and outlook} \setcounter{equation}{0}

In this paper we have performed a general study on the conditions under which locally stable 
vacua with spontaneously broken supersymmetry can occur in ${\cal N}\!=2$ supergravity theories 
with only hypermultiplets. The results have been compared with the corresponding conditions that 
were already known for ${\cal N}\!=1$ supergravity theories with only chiral multiplets \cite{GRS,CGGLPS}. As in 
the ${\cal N}\!=1$ case, our strategy has been to look at the most dangerous scalar fluctuations, 
which are the ones related to the scalar partners of the Goldstino fermion, the sGoldstinos. 

In the ${\cal N}\!=1$ case the constraint can be formulated as a lower bound on the curvature of the 
scalar manifold spanned by the scalar components of the chiral multiplets. More concretely, they 
represent a lower bound on the holomorphic sectional curvature in the complex sGoldstino direction
defined by the complex structure of the K\"ahler-Hodge scalar manifold. 
They constrain therefore both the allowed scalar geometries and the allowed 
supersymmetry breaking directions. In the ${\cal N}\!=2$ case, we have found that there are 
three constraints on the curvature of the scalar manifold, which are all similar to the one arising 
in ${\cal N}\!=1$ theories. This corresponds to the fact that in this case there are more sGoldstinos. 
More precisely, one finds an upper bound on the three possible 
holomorphic sectional curvatures in the complex sGoldstino directions defined by the three almost 
complex structures of the quaternionic-K\"ahler scalar manifold.
However, it turns out that the quaternionic-K\"ahler geometry underlying ${\cal N}\!=2$ models 
implies a very restricted form of the curvature tensor, which is completely fixed up to a Weyl-type 
contribution $\Sigma$. This is in contrast with the K\"ahler-Hodge geometry underlying ${\cal N}\!=1$ 
theories, which allows instead for a generic curvature tensor. As a consequence, the average of 
the three holomorphic sectional curvatures arising in the ${\cal N}\!=2$ constraints happens to 
have a fixed constant value independent of $\Sigma$, which translates into a universal negative 
upper bound on the values of the cosmological constant that are compatible with the metastability 
of the vacuum. This implies in particular that metastable dS vacua are excluded, independently 
of the specific scalar geometry of the model.\footnote{Under certain (restrictive) circumstances, 
it is possible to consistently truncate an ${\cal N}\!=2$ theory with $n$ hypermultiplets down to an 
${\cal N}\!=1$ theory with $n'$ chiral multiplets \cite{ADF}. At the geometrical level, this truncation 
involves the restriction to a K\"ahler-Hodge submanifold of the original quaternionic-K\"ahler 
manifold. Even though the curvature of the  K\"ahler submanifold is
arbitrary, the superpotential and thus the sGoldstino directions are
more constraint than in generic ${\cal N}\!=1$ theories.
It would be interesting to study in  more detail the stability
conditions in this case.}

The strong results that we find for ${\cal N}\!=2$ theories in the case with only hypermultiplets are 
very similar to the comparably strong results holding in the case in which 
only vector multiplets are present and the gauging is Abelian \cite{DW,CKVDFDG}. 
They both have to do with the restricted form that the curvature of the 
scalar manifolds, which are respectively quaternionic-K\"ahler and special-K\"ahler, is allowed
to take. In fact, the upper bounds on the lowest mass eigenvalue in these two 
special cases read
\be\begin{aligned}
m^2_{\rm hyper} &\le - V - \tfrac13 |m_{3/2}|^2\,, \label{c1} \\
m^2_{\rm vector} &\le - 2 V\,.
\end{aligned}\ee
Similar tachyonic modes seem to be endemic also in ${\cal N}\! > 2$ theories; see for instance 
refs.~\cite{KLPS}.

Another interesting information one can deduce from the stability bounds (\ref{c1})
concerns dS stationary points. For example they could be of potential interest 
for achieving inflation. Note nevertheless that there will be at least one direction in field space along 
which $|V''|/V \sim 1$, implying that the conditions for slow-roll inflation are never satisfied.

In more general situations of ${\cal N}\!=2$ supergravity theories involving both vector and 
hypermultiplets, as well as non-Abelian gauging, some examples of models giving rise to 
dS spaces are known to exist \cite{FTV}. It is clear that an analysis 
of the same type as the one presented here for these more general situations would also be 
very valuable, as it could provide some insights on what are the really necessary ingredients 
to construct models admitting a stable dS vacuum \cite{DGLS}. For instance, it is obvious 
that non-Abelian gaugings help, since  then a new positive-definite term arises in 
the scalar potential. But even for Abelian gaugings, combining vector multiplets with 
hypermultiplets may be sufficient to be able to avoid tachyons, since in that case the 
scalar manifold is the product of a quaternionic-K\"ahler and special-K\"ahler manifolds, which as a 
whole can have a lower sectional curvature than any of its two components. Of course, even after having 
understood more precisely the conditions for achieving dS vacua within ${\cal N}\!=2$ supergravity 
effective theories, another interesting question would be whether these can be realized in 
string theory.

\section*{Acknowledgments}

This work was partly supported by the German Science Foundation (DFG)
under the Collaborative Research Center (SFB) 676, by the European Union 
6th Framework Program MRTN-CT-503369 ``Quest for
unification" and by the Swiss National Science Foundation. 

We would like to thank G.~Dall'Agata for several crucial discussions. We are also 
grateful to N.~Ambrosetti, L.~Andrianopoli, J.~P.~Derendinger, S.~Ferrara,
J.~Gauntlett, F. Saueressig, G.~Villadoro  and M.~Zagermann for useful
discussions.  

J.L. thanks Luis Alvarez-Gaum\'e and the CERN Theory Division for
hospitality and financial support during the initial part of this work.

\vskip 1cm

\appendix

\renewcommand{\theequation}{\thesection.\arabic{equation}}
\setcounter{equation}{0}

\noindent {\bf \Large Appendix} 

\section{Supertrace sum rule on the masses} \setcounter{equation}{0}

In this Appendix we report some details on the computation of the supertrace of the 
square mass operator for all the fields. This quantity is of some interest, since it controls 
the leading quadratic divergences arising at the one loop level when supersymmetry is 
spontaneously broken, at least in the case of a flat Minkowski space with vanishing
cosmological constant. We will first shortly review the know case of ${\cal N}\!=1$ 
theories and then present the same computation for ${\cal N}\!=2$ models.

\subsection{${\cal N}\!=1$ theories with chiral multiplets}\label{sumruleN1}

Using the expressions given in Section 2.1 for the mass matrices of the various fields, one finds that at 
a generic stationary point with any allowed cosmological constant:
\bea
{\rm tr} [m_0^2] &=& 2 \nabla_i N_k \nabla^i \bar N^k 
- 2 R_{i \bar \jmath} \bar N^i N^{\bar \jmath} + 2 (n-1) \bar N^k N_k - 4 n |L|^2 \,, \\
{\rm tr} [m_{1/2}^2] &=& \nabla_i N_k \nabla^i \bar N^k - \tfrac {8}{3} \bar N^k N_k 
+ \tfrac {4}{9} (\bar N^k N_k)^2 |L|^{-2} \,, \\
{\rm tr} [m_{3/2}^2] &=& |L|^2 \,.
\eea
It follows that: \cite{GRK}
\be\begin{aligned}
{\rm str} [m^2] &= {\rm tr} [m_0^2] - 2\, {\rm tr} [m_{1/2}^2] - 4\,  {\rm tr} [m_{3/2}^2]  \\
&= 2 (n-1) m_{3/2}^2 + 2 R_{i \bar \jmath} \bar N^i N^{\bar \jmath} + 2 (n-1) V - \tfrac 89 V^2 |m_{3/2}|^{-2} \,.
\end{aligned}\ee
In terms of $\gamma = V/(3|m_{3/2}|^2)$, this finally gives:
\be
{\rm str} [m^2] = \big[2 (n-1) + 6 (n-1) \gamma - 8 \gamma^2 \big] |m_{3/2}|^2 
+ 2 R_{i \bar \jmath} \bar N^i N^{\bar \jmath} \,. 
\ee

Note that for K\"ahler manifolds that happen to be also Einstein spaces, with a Ricci tensor of the form
\be
R_{i \bar \jmath} = r g_{i \bar \jmath} \,,
\ee
the formula simplifies as follows: 
\be
{\rm str} [m^2] = \big[2 (n-1+3 r) + 6 (n-1+r) \gamma - 8 \gamma^2 \big] |m_{3/2}|^2\,.
\ee
Note finally that for supersymmetric vacua with $\gamma=-1$ one finds:
\be
{\rm str} [m^2] = - 4 (n+1) |m_{3/2}|^2 \,.
\ee

\subsection{${\cal N}\!=2$ theories with hypermultiplets}

Using the expressions derived in Section 3.1 for the mass matrices of the various fields, 
as well as eq.~(\ref{Ricci}), one can compute the traces of the square mass for each field
at a generic stationary point of the scalar potential. After some algebra, and repeated 
use of the various identities listed at the beginning of Section 3, we find the following results:
\bea
{\rm tr} [m_0^2] 
&=& 4 \nabla^u k^v \nabla_u k_v + 4 (2n -5) k^u k_u - 12 n P^x P^x \,, \\
{\rm tr} [m_{1/2}^2] 
&=& 2 \nabla^u k^v \nabla_u k_v - 16 k^u k_u - 2 n P^x P^x + \tfrac {128}{9} (k^u k_u)^2 (P^x P^x)^{-1}\,, \\
{\rm tr} [m_1^2] 
&=&4 k^u k_u \,, \\[0.5mm]
{\rm tr} [m_{3/2}^2]
&=& 2 P^x P^x \,.
\eea
Using these result, the supertrace is found to be:
\be\begin{aligned}
{\rm str} [m^2] &= {\rm tr} [m_0^2] - 2\, {\rm tr} [m_{1/2}^2] + 3\, {\rm tr} [m_1^2] - 4\, {\rm tr} [m_{3/2}^2]  \\
&= - \big(2 n + 6\big)\, |m_{3/2}|^2 + \big(2 n - \tfrac {14}{3}\big)\, V - \tfrac {16}9\, V^2 |m_{3/2}|^{-2} \,.
\end{aligned}\ee
In terms of $\gamma = V/(3|m_{3/2}|^2)$, this finally reads:
\be
{\rm str} [m^2] = \big[\!-\! \big(2 n + 6\big) + \big(6 n - 14\big) \gamma - 16 \gamma^2 \big] |m_{3/2}|^2 \,. 
\ee
Note that for supersymmetric vacua with $\gamma = -1$ one finds:
\be
{\rm str} [m^2] = - 8 (n + 1)\, |m_{3/2}|^2 \,.
\ee

\section{Curvature conventions} \setcounter{equation}{0}

In this Appendix, we summarize our conventions for the curvature tensor
and the sectional curvature, first for generic real Riemann manifolds and 
then for complex K\"ahler manifolds.

\subsection{Riemann manifolds}

For the geometry of a generic real Riemann manifold, we use the following 
conventions. Denoting the components of the metric with $g_{uv}$, the 
Christoffel connection is $\Gamma^{k}_{uv} = \frac 12 g^{k l} 
\big(\partial_u g_{vl} + \partial_v g_{ul} - \partial_l g_{uv} \big)$.
The Riemann tensor is defined as
\be
R^u_{\;vkl} = \partial_k \Gamma^u_{vl} - \partial_l \Gamma^u_{vk} 
+ \Gamma^i_{ks} \Gamma^s_{jl} - \Gamma^i_{ls} \Gamma^s_{jk} \,. 
\ee
The Ricci curvature tensor is then:
\be
R_{i j} = R^s_{\;isj} \,,
\ee
and finally the scalar curvature is given by:
\be
R = R^s_{\;s} \,.
\ee
The ordinary covariant derivatives on vectors is defined as 
${\cal D}_u V_v = \partial_u V_v - \Gamma^s_{uv} V_s$,
and the commutator of two of them gives:
\be
\big[{\cal D}_u, {\cal D}_v \big] V_k = R_{u v k}^{\;\;\;\;\;\;l} V_l \,.
\ee
The sectional curvature in a plane defined by two orthogonal vectors $A_u$ and $B_v$,
with $A^u B_u = 0$, is finally defined as:
\be
R(A,B) = \frac {R_{uvkl} A^u B^v A^k B^l}{A^r A_r \, B^s B_s} \,.
\ee

\subsection{K\"ahler manifolds}

For complex K\"ahler manifolds admitting a globally-defined complex structure 
$J_{uv}$, it is convenient to switch to complex coordinates in which this is block 
diagonal with values $\pm i$. The Hermitian metric has non-vanishing 
components $g_{i \bar \jmath}$ and $g_{\bar \imath j}$, and satisfies the 
conditions $\partial_i g_{j \bar k} = \partial_j g_{i \bar k}$ and 
$\partial_{\bar \imath} g_{\bar \jmath k} = \partial_{\bar \jmath} g_{\bar \imath k}$. 
It follows then that the non-vanishing components of the Christoffel 
connection are $\Gamma^k_{ij} = g^{k \bar l} \partial_{i} g_{j \bar l}$ and 
$\Gamma^{\bar k}_{\bar \imath \bar \jmath} = g^{\bar k l} \partial_{\bar \imath} g_{\bar \jmath l}$.
The non-vanishing components of the Riemann tensor are then:
\bea
R_{i \bar \jmath k \bar l} &=& \partial_i \partial_{\bar \jmath} g_{k \bar l} 
+ g^{\bar r s} \partial_i g_{k \bar r} \partial_{\bar \jmath} g_{\bar l s} \,, \\
R_{\bar \imath j k \bar l} &=& - R_{j \bar \imath k \bar l} \,,\;\;
R_{i \bar \jmath \bar k l} = - R_{i \bar \jmath l \bar k} \,,\;\;
R_{\bar \imath j \bar k l} = R_{j \bar \imath l \bar k} \,. 
\eea
The Riemann tensor has in this case the additional property of being symmetric under 
the exchange of indices of the same holomophic or antiholomorphic type: 
$R_{i \bar \jmath k \bar l} = R_{k \bar \jmath i \bar l} = R_{i \bar l k \bar \jmath} = R_{k \bar l i \bar \jmath}$. 
The Ricci curvature tensor has then as only non-vanishing components
\bea
R_{i \bar \jmath} = - g^{r \bar s} R_{r \bar s i \bar \jmath} \,, \qquad
R_{\bar \imath j} = R_{j \bar \imath}\,.
\eea
Finally, the scalar curvature is given by:
\be
R = 2g^{r \bar s} R_{r \bar s} \,.
\ee
The ordinary covariant derivatives on holomorphic vectors (similar formulae hold for antiholomorphic 
vectors) read ${\cal D}_i V_j = \partial_i V_j - \Gamma^s_{ij} V_s$ and ${\cal D}_{\bar \imath} V_j = \partial_{\bar \imath} V_j$, 
and the commutator of two of them gives:
\be
\big[{\cal D}_i, {\cal D}_{\bar \jmath} \big] V_k = R_{i \bar \jmath k}^{\;\;\;\;\; l} V_l \,.
\ee
The holomorphic sectional curvature in a plane defined by a vector and its conjugate under
the complex structure, defining in complex coordinates a holomorphic vector $Z_i$ and its 
antiholomorphic counterpart $Z_{\bar \imath}$, is finally given by:
\be
R(Z) = - \frac {R_{i \bar \jmath k \bar l} Z^i Z^{\bar \jmath} Z^k Z^{\bar l}}{(Z^p Z_p)^2} \,.
\ee

\end{document}